# On the motion of passive and active particles with harmonic and viscous forces

Jae-Won Jung[1], Sung Kyu Seo[1], Kyungsik Kim[1,2]

[1]*DigiQuay Ltd., Manan-Gu Anyang, Gyeongggido 14084, Republic of Korea*
[2]*Department of Physics, Pukyung National University, Busan 608-737, Korea*

In this paper, we solve the joint probability density for the passive and active particles with harmonic, viscous, and perturbative forces. After deriving the Fokker-Planck equation for a passive and a run-and-tumble particles, we approximately get and analyze the solution for the joint distribution density subject to an exponential correlated Gaussian force in three kinds of time limit domains. Mean squared displacement (velocity) for a particle with harmonic and viscous forces behaviors in the form of super-diffusion scales as $\sim t^4$ ( $\sim t^2$ ) in $\tau = 0$ ( $t << \tau$ ), consistent with a particle having viscous and perturbative forces. A passive particle with both harmonic, viscous forces and viscous, perturbative forces has the Gaussian form with mean squared velocity $<v^2(t)> \sim t$ in $\tau = 0$. Particularly, In our case of a run-and-tumble particle, the mean squared displacement scales to $<x_\pm^2(t)> \sim t^3$ in $t << \tau$, while the mean squared velocity has a normal diffusive form with $<v_\pm^2(t)> \sim t$ in $t >> \tau$ and $\tau = 0$. In addition, the kurtosis, the correlation coefficient, and the moment from moment equation are numerically calculated.

-----------------------------------------------------------------------------------
* Corresponding authors. E-mail: kskim@pknu.ac.kr

# I. Introduction

As studied and analyzed by some researchers in the past time, the dynamical behavior of passive particles through the thermal contact between the system and the heat reservoir [1-3], which is the crucial subject of statistical mechanics [4-7], has been mainly examined and discussed by solving the generalized Langevin equation and the Fokker-Planck equation [8]. The motion of interacting and correlating particles is controlled by the energy of the particle, the resistance of the motion, the thermal noise, etc., and is supplemented by the energy flowing into the outside heat reservoirs or other systems. Sub-diffusive, super-diffusive, and ballistic behaviors [9, 10] for displacement and velocity have already been addressed in diffusion and transport problems, and unproven dynamic behaviors are emerging along with experiments. The problems of solving the force equation, generalized Langevin equation, with harmonic, perturbative, and other forces, have been utilized by many scientists in diverse fields. In addition, such problems in several sorts of mediums include the statistical values of various diffusion constants, first passage times, correlation coefficients, memory functions, and so on. The fractional Brownian motion [11], fractional diffusion, and Fokker-Planck equation [12-15] have treated and analyzed theoretically and numerically in the anomalous diffusion process and the transport process. Active Brownian particle recently studied its behavior via the equation of motion within a harmonic trap in the presence of translational diffusion, and they presented explicit calculation of several moments at arbitrary times and their evolution to the steady state [16]. In addition, the steady state velocity distribution has exhibited a re-entrant crossover from passive Gaussian to active non-Gaussian behaviors and constructed a corresponding the phase diagram using the exact expression of the kurtosis [17].

As is well known in stochastic processes, there have been theoretically studied over many years on the Brownian motion described the statistical properties of classical and modern systems through the generalized Langevin equation extended to the fluctuation-dissipation theorem. Interacting and correlating particles have currently been classified into two kinds of particles: the passive particle and the active particle. That is, the former moves without position it wants like Brownian particle, the latter moves to the position it wants with a sense of purpose like micro-swimmer. Micro-active particles are excavated experimentally, and are underway novel research. The collective and coherent motion for large numbers of self-propelled organisms has studied and simulated from the long-term ticked data.

There have recently been variously and satisfactorily published papers about computer-simulations and experimental devices with the utility, the generality, and the precision from the Brownian particle, colloid, run-and-tumble particle [18-21], micro-swimmer to flock, herd, swarm, slime models.

Through computer-simulations [22-25] and experimental devices, the features between passive and active particles have been described and analyzed by different fluctuations and statistical properties [26, 27]. Examples of active particle systems include Brownian motors [28], motile cells [29-31], microscopic animals [32, 33], self-propelled particles [34, 35], interacting collective particles [36-38], automated digital tracking [39-41], and active granular colloidal systems [42-46].

However, the dynamical behavior of a passive or active particle may be easily and extensively understood and analyzed its calculated moment via Fourier, Laplace transform in the diffusion and transport process if the probability density is not obtained or solved by the equation of its motion. In this paper, approximately the solution of the joint probability density having two variables of displacement and velocity is approximately solved in the limits of $t \ll \tau, t \gg \tau$, and for $\tau = 0$, where $t \ll \tau$ is the correlation time. The paper is structured as follows: In section 2, after deriving the Fokker-Planck equation, we obtain the joint probability density for a passive particle with harmonic and viscous forces, subject to an exponentially correlated Gaussian force. In section 3, we obtain the joint probability density for a run-and-tumble particle by similar method as solved of section 2. In addition, the kurtosis, the correlation coefficient, and the moment from moment equation are numerically calculated. Finally, we provide a conclusion summarizing our key findings in section 4.

## II. Fokker-Planck equation in harmonic force

### 2-1 Fokker-Planck equation for joint probability distribution function in displacement and velocity

First of all two coupled equations for a passive particle having the displacement and the velocity are expressed in terms of

$$\frac{dx}{dt} = v(t) \qquad (2\text{-}1)$$

$$\frac{dv(t)}{dt} = -rv(t) - \beta x(t) + f(t). \qquad (2\text{-}2)$$

Here, $-rv(t)$ denotes the viscous force consuming on the particles, $r$ the viscous coefficient, and $-\beta x(t)$ the harmonic force. The value $g(t)$ is the fluctuating force activated by the surrounding small particles by the fluctuation of the particle. In Eq. (2-1) and Eq. (2-2), the fluctuation term of an exponential correlated Gaussian force is given by $<f(t)> = 0$, $<f(t)f(t')> = f_0^2 f(t-t')$ and $f(t-t') = \frac{1}{2\tau}\exp(-\frac{|t-t'|}{\tau})$ for $f_0^2 = 2rk_B T_{f_0}$. The joint probability density $P(x(t), v(t), t)$ for the displacement $x$ and the velocity $v$ at time $t = 0$ is defined by $P(x(t), v(t), t) = <\delta(x-x(t))\delta(v-v(t))>$. It is assumed that the particle is at rest at $x = 0$, $v = 0$ in initial state. By taking time derivatives of the joint probability density, we have

$$\frac{\partial P(x,v,t)}{\partial t} = -\frac{\partial}{\partial x}<\frac{\partial x}{\partial t}\delta(x-x(t))\delta(v-v(t))> - \frac{\partial}{\partial v}<\frac{\partial v}{\partial t}\delta(x-x(t))\delta(v-v(t))>. \qquad (2\text{-}3)$$

Inserting Eq. (2-1) and Eq. (2-2) in Eq. (2-3), we can write

$$\frac{\partial P(x,v,t)}{\partial t} = -v\frac{\partial P(x,v,t)}{\partial x} + <[rv(t) + \beta x(t) - f(t)]\delta(x-x(t))\delta(v-v(t))>. \qquad (2\text{-}4)$$

By manipulating over integrals using Eq. (2-4) [47], the joint probability density $P(x(t), v(t), t)$ is derived as

$$\frac{\partial P(x,v,t)}{\partial t} = -v\frac{\partial P(x,v,t)}{\partial x} + r\frac{\partial}{\partial v}vP(x,v,t) + \beta x\frac{\partial P(x,v,t)}{\partial v}$$
$$-ac(t)\frac{\partial^2 P(x,v,t)}{\partial x \partial v} + ab(t)\frac{\partial^2 P(x,v,t)}{\partial v^2} \qquad (2\text{-}5)$$

Here $a = f_0^2/2$, $c(t) = (t+\tau)\exp(-t/\tau) - \tau$ and $b(t) = 1 - \exp(-t/\tau)$.

## 2-2 $P(x,t)$ and $P(v,t)$ in short-time domain

In order to find the joint probability density, we introduce the double Fourier transform

$$P(\xi,\eta,t) = \int_{-\infty}^{+\infty} dx \int_{-\infty}^{+\infty} dt \exp(-i\xi x - i\eta v) P(x,v,t). \qquad (2\text{-}6)$$

By taking the double Fourier transform in Eq. (2-5), Fourier transform of joint probability density is given by

$$\frac{\partial}{\partial t}P(\xi,\eta,t) = (\xi - r\eta)\frac{\partial}{\partial \eta}P(\xi,\eta,t) - \beta\eta\frac{\partial}{\partial \xi}P(\xi,\eta,t) + a[c(t)\xi\eta - b(t)\eta^2]P(\xi,\eta,t). \qquad (2\text{-}7)$$

In steady state, introducing $\frac{\partial}{\partial t}P(\xi,\eta,t) = 0$ and $P(\xi,\eta,t) \to P^{st}(\xi,\eta,t)$, then we can calculate as

$$(\xi - r\eta)\frac{\partial}{\partial \eta}P^{st}(\xi,\eta,t) - \beta\eta\frac{\partial}{\partial \xi}P^{st}(\xi,\eta,t) + a[c(t)\xi\eta - b(t)\eta^2]P^{st}(\xi,\eta,t) = 0. \qquad (2\text{-}8)$$

Two equations from $P(\xi,\eta,t) = P_\xi^{st}(\xi,t)P_\eta^{st}(\eta,t)$ in respect with obtaining the special solutions for $\xi$, $\eta$ by variable separation is as follows:

$$-\beta\eta\frac{\partial}{\partial \xi}P_\xi^{st}(\xi,t) + \frac{1}{2}a[c(t)\xi\eta - b(t)\eta^2]P_\xi^{st}(\xi,t) + AP_\xi^{st}(\xi,t) = 0 \qquad (2\text{-}9)$$

$$(\xi - r\eta)\frac{\partial}{\partial \eta}P_\eta^{st}(\eta,t) + \frac{1}{2}a[c(t)\xi\eta - b(t)\eta^2]P_\eta^{st}(\eta,t) - AP_\eta^{st}(\eta,t) = 0. \qquad (2\text{-}10)$$

From Eq. (2-9), Fourier transform of the probability density for the velocity in the steady state is obtained as

$$P_\xi^{st}(\xi,t) = \exp[\frac{a}{2\beta\eta}[\frac{c(t)}{2}\eta\xi^2 - b(t)\eta^2\xi]] \qquad (2\text{-}11)$$

So to speak, in order to find the solution of joint function for $\xi$ from $P_\xi(\xi,t) \equiv Q_\eta(\eta,t)P_\eta^{st}(\eta,t)$, we can obtain the distribution functions in the short-time domain $t \ll \tau$ via the calculation including terms up to order $t^2/\tau^2$. That is,

$$P_\xi(\xi,t) = Q_\xi(\xi,t)\exp[\frac{a}{2\beta\eta}[\frac{c(t)}{2}\eta\xi^2 - b(t)\eta^2\xi]] \qquad (2\text{-}12)$$

$$Q_\xi(\xi,t) = R_\xi(\xi,t)\exp[\frac{a}{2(\beta\eta)^2}[\frac{b'(t)}{2}\eta^2\xi^2 - \frac{c'(t)}{6}\eta\xi^3]] \qquad (2\text{-}13)$$

$$R_\xi(\xi,t) = S_\xi(\xi,t)\exp[-\frac{a}{2(\beta\eta)^3}[\frac{b''(t)}{6}\eta^2\xi^3 - \frac{c''(t)}{24}\eta\xi^4]] \qquad (2\text{-}14)$$

$$S_\xi(\xi,t) = T_\xi(\xi,t)\exp[\frac{a}{2(\beta\eta)^4}[\frac{b'''(t)}{24}\eta^2\xi^4 - \frac{c'''(t)}{120}\eta\xi^5]]. \qquad (2\text{-}15)$$

Here, neglecting terms proportional to $1/\tau^3$ and taking the solutions as arbitrary functions of variable $t-(\xi/\beta\eta)$, the arbitrary function $T_\xi(\xi,t)$ becomes $T_\xi(\xi,t)=\Theta_\xi[t-(\xi/\beta\eta)]$. We consequently derive that

$$P_\xi(\xi,t) = \Theta[t-(\xi/\beta\eta)]S_\xi^{st}(\xi,t)R_\xi^{st}(\xi,t)Q_\xi^{st}(\xi,t)P_\xi^{st}(\xi,t). \tag{2-16}$$

By expanding their derivatives to second order in powers of $t/\tau$, we obtain the expression for $P_\xi(\xi,t)$ after some cancellations. That is,

$$P_\xi(\xi,t) = \exp[-\frac{at^3}{12\beta\tau^2}[1-3[\frac{t}{\tau}]^{-1}]\xi^2 - \frac{at^3}{4\tau}[1-\frac{1}{6}[\frac{t}{\tau}]]\eta\xi - \frac{a\beta t^3}{4}[1-\frac{1}{3}[\frac{t}{\tau}]]\eta^2]. \tag{2-17}$$

Here,

$$\Theta_\xi[u] = \exp[-\frac{a\beta}{120\tau^2}\eta^2 u^5 + \frac{a\beta}{48\tau}\eta^2 u^4 + \frac{a\beta t}{12\tau}\eta^2 u^3 - \frac{a}{4\tau}\eta^2 u^2 - \frac{a\beta t}{4}\eta^2 u^2]. \tag{2-18}$$

Using the inverse Fourier transform, the probability density $P(x,t)$ is presented by

$$P(x,t) = \frac{1}{2\pi}\int_{-\infty}^{+\infty} d\xi \exp(-i\xi x)P_\xi(\xi,t)$$

$$= [\pi\frac{at^3}{3\beta\tau^2}[1-3[\frac{t}{\tau}]^{-1}]]^{-1/2}\exp[-\frac{3\beta\tau^2 x^2}{at^3}[1-3[\frac{t}{\tau}]^{-1}]]. \tag{2-19}$$

The mean squared displacement for $P(x,t)$ is given by

$$<x^2> = \frac{at^3}{6\beta\tau^2}[1-3[\frac{t}{\tau}]^{-1}]. \tag{2-20}$$

For obtaining the special solutions for $\eta$ in the short-time domain $t\ll\tau$, we use the variable-separated equation like Eq. (2-10). In the steady state, as we assume that $\frac{1}{(\xi-r\eta)}\cong\frac{1}{\xi}(1+\frac{r\eta}{\xi})$, the probability density becomes

$$P_\eta^{st}(\eta,t) = \exp[-\frac{a}{2\xi}[-\frac{b(t)}{3}\eta^3+\frac{c(t)}{2}\xi\eta^2]+\frac{ar}{2\xi^2}[\frac{b(t)}{4}\eta^4-\frac{c(t)}{3}\xi\eta^3]] \tag{2-21}$$

Through the similar method, Fourier transform of the probability density $P_\eta(\eta,t)=Q_\eta(\eta,t)P_\eta^{st}(\eta,t)$ and other ones are obtained by

$$P_\eta(\eta,t) = Q_\eta(\eta,t)\exp[-\frac{a}{2\xi}[-\frac{b(t)}{3}\eta^3+\frac{c(t)}{2}\xi\eta^2]+\frac{ar}{2\xi^2}[\frac{b(t)}{4}\eta^4-\frac{c(t)}{3}\xi\eta^3]] \tag{2-22}$$

$$Q_\eta(\eta,t) = R_\eta(\eta,t)\exp[\frac{a}{2\xi^2}[\frac{b'(t)}{12}\eta^4-\frac{c'(t)}{6}\xi\eta^3]+\frac{ar}{2\xi^2}[\frac{b'(t)}{4}\eta^5-\frac{c'(t)}{3}\xi\eta^4]] \tag{2-23}$$

$$R_\eta(\eta,t) = S_\eta(\eta,t)\exp[\frac{a}{2\xi^3}[\frac{b''(t)}{60}\eta^5-\frac{c''(t)}{24}\xi\eta^4]+\frac{ar}{2\xi^4}[\frac{b''(t)}{120}\eta^6-\frac{c''(t)}{60}\xi\eta^6]] \tag{2-24}$$

$$S_\eta(\eta,t) = T_\eta(\eta,t)\exp[-\frac{a}{\xi^3}\frac{c'''(t)}{240}\eta^5-\frac{ar}{\xi^4}\frac{c'''(t)}{720}\eta^6]. \tag{2-25}$$

Here, $b'(t)=\frac{db(t)}{dt}$ and $c'(t)=\frac{dc(t)}{dt}$. Discarding terms proportional to $1/\tau^3$, $T_\eta(\eta,t)$ is given by

$$T_\eta(\eta,t) = (\xi-r\eta)\frac{\partial}{\partial\eta}T_\eta(\eta,t). \tag{2-26}$$

Taking the solutions as arbitrary functions of variable $t+\eta/(\xi-r\eta)$, then the arbitrary function becomes $T_\eta(\eta,t) = \Theta_\eta[t + [\eta/(\xi - r\eta)]]$. By expanding their derivatives to second order in powers of $t/\tau$, we also obtain the expression for $P_\eta(\eta,t)$ after some cancellations, that is,

$$P_\eta(\eta,t) = \Theta_\eta[t + \eta/(\xi - r\eta)]S_\eta^{st}(\eta,t)R_\eta^{st}(\eta,t)Q_\eta^{st}(\eta,t)P_\eta^{st}(\eta,t)$$

$$= \exp[-\frac{5a}{4}\frac{t^2}{\tau}[1-\frac{2}{3}rt]\eta^2 - \frac{a}{2}\frac{t^3}{\tau}[1-\frac{5}{3}rt]\xi\eta - \frac{a}{8}\frac{t^4}{\tau}[1-2[\frac{t}{\tau}]^{-1}]\xi^2]. \quad (2\text{-}27)$$

Here,

$$\Theta_\eta[u] = \exp[-\frac{a}{48\tau}\xi^{-2}(\xi-rv)^4 u^4 - \frac{ar}{120\tau}\xi^{-3}(\xi-r\eta)^5 u^5 - \frac{a}{24\tau}\xi^{-2}(\xi-r\eta)^4 u^4$$

$$+\frac{at}{12\tau}\xi^{-1}(\xi-r\eta)^3 u^3 + \frac{ar}{40\tau}\xi^{-3}(\xi-r\eta)^5 u^5 + \frac{art}{24\tau}\xi^{-2}(\xi-r\eta)^4 u^4$$

$$-\frac{at}{6\tau}\xi^{-1}(\xi-r\eta)^3 u^3 + \frac{at}{4}(\xi-r\eta)^2 u^5 + \frac{art}{8\tau}\xi^{-2}(\xi-r\eta)^4 u^4] \quad . \quad (2\text{-}28)$$

Using the inverse Fourier transform, the probability density $P(v,t)$ is presented by

$$P(v,t) = \frac{1}{2\pi}\int_{-\infty}^{+\infty} d\eta \exp(-i\eta v)P_\eta(\eta,t) = [\pi\frac{5a}{2}\frac{t^2}{\tau}[1-\frac{2}{3}rt]]^{-1/2}\exp[-\frac{2\tau v^2}{5at^3}[1-\frac{2}{3}rt]^{-1}] \quad (2\text{-}29)$$

The mean squared velocity for $P(v,t)$ is given by

$$<v^2> = \frac{5a}{4}\frac{t^2}{\tau}[1-\frac{2}{3}rt]. \quad (2\text{-}30)$$

## 2-3 $P(x,t)$ and $P(v,t)$ in long-time domain

In long-time domain, we write the approximate equation for $x$

$$\frac{\partial}{\partial t}P_\xi(\xi,t) \cong \frac{a}{2}[c(t)\xi\eta - b(t)\eta^2]P_\xi(\xi,t). \quad (2\text{-}31)$$

Then we can simply calculate $P^{st}(\xi,t)$ as

$$P_\xi(\xi,t) = \exp[\frac{a}{2}\int[c(t)\xi\eta - b(t)\eta^2]\, dt. \quad (2\text{-}32)$$

As $\int b(t)dt = t-\tau$, $b(t) = 1$ and $\int c(t)dt = -\tau t$, $c(t) = -\tau$ in long-time domain, we can find $Q_\xi^{st}(\xi,t)$ for $\xi$ from $P_\xi(\xi,t) \equiv Q_\xi(\xi,t)P_\xi^{st}(\xi,t)$, that is,

$$Q_\xi^{st}(\xi,t) = \exp[\frac{a}{2}\int[b(t)\eta^2 - c(t)\xi\eta]\, dt]. \quad (2\text{-}33)$$

From Eq. (2-9), we have for long-time domain $t \gg \tau$

$$P_\xi^{st}(\xi,t) = \exp[\frac{a}{2\beta\eta}[\frac{c(t)}{2}\eta\xi^2 - b(t)\eta^2\xi]] \quad (2\text{-}34)$$

$$Q_\xi(\xi,t) = R_\xi(\xi,t)Q_\xi^{st}(\xi,t) = R_\xi(\xi,t)\exp[\frac{a}{2}\int[b(t)\eta^2 - c(t)\eta\xi]\, dt]. \quad (2\text{-}35)$$

Taking the solutions as arbitrary functions of variable $t - \xi/\beta\eta$, the arbitrary function $R_\xi(\xi,t)$ becomes $R_\xi(\xi,t) = \Theta_\xi[t - [\xi/\beta\eta]]$. Consequently, as expanding their derivatives to second order in powers of $t/\tau$, we respectively obtain the bilinear expressions for $P_\xi(\xi,t)$, after some cancellations. That is,

$$P_\xi(\xi,t) = R_\xi(\xi,t)Q_\xi^{st}(\xi,t)P_\xi^{st}(\xi,t) = \Theta_\xi[t-[\xi/\beta\eta]])Q_\xi^{st}(\xi,t)P_\xi^{st}(\xi,t) \ . \quad (2\text{-}36)$$

Here,

$$\Theta_\xi(u) = \exp[\frac{a}{2}\beta\tau t\eta^2 u - \frac{a}{2}(t-\tau)\eta^2 + \frac{a\beta\tau}{4}v^2 u^2 - \frac{a}{2}\eta^2 u] . \quad (2\text{-}37)$$

In long-time domain, we write approximate equation for $\eta$

$$\frac{\partial}{\partial t}P_\eta(\eta,t) \cong \frac{a}{2}[c(t)\xi\eta - b(t)\eta^2]P_\eta(\eta,t) \ . \quad (2\text{-}38)$$

We simply calculate $P_\eta(\eta,t)$ as

$$P_\eta(\eta,t) = \exp[\frac{a}{2}\int[c(t)\xi\eta - b(t)\eta^2]\,dt \ . \quad (2\text{-}39)$$

We can find the solutions of functions for $\xi$ from $Q_\eta(\eta,t)P_\eta(\eta,t)$, That is,

$$Q_\eta^{st}(\eta,t) = \exp[\frac{a}{2}\int[b(t)\eta^2 - c(t)\xi\eta]\,dt \ . \quad (2\text{-}40)$$

We calculate for long-time domain $t \gg \tau$

$$P_\eta^{st}(\eta,t) = \exp[-\frac{a}{2\xi}[-\frac{b(t)}{3}\eta^3 + \frac{c(t)}{2}\xi\eta^2] + \frac{ar}{2\xi^2}[\frac{b(t)}{4}\eta^4 - \frac{c(t)}{3}\xi\eta^3]] \quad (2\text{-}41)$$

It is clear in long-time domain that Fourier transform of probability density $P^{st}(v,t)$ for the velocity $v$ and Eq. (2-41) should be the same. Taking the solutions as arbitrary functions of variable $t+\eta/[\xi-r\eta]$, the arbitrary function $R_\eta(\eta,t)$ becomes $R_\eta(\eta,t) = \Theta_\eta[t+\eta/[\xi-r\eta]]$. Consequently, as expanding their derivatives to second order in powers of $t/\tau$, we respectively obtain the expression for $P_\eta(\eta,t)$ after some cancellations. That is,

$$P_\eta(\eta,t) = R_\eta(\eta,t)Q_\eta^{st}(\eta,t)P_\eta^{st}(\eta,t)$$
$$= \Theta_\eta[t+\eta/[\xi-r\eta]]Q_\eta^{st}(\eta,t)P_\eta^{st}(\eta,t) \quad (2\text{-}42)$$

Therefore, from Eq. (2-36) and Eq. (2-42), Fourier transform of joint probability density $P(x,v,t)$ is calculated as

$$P(\xi,\eta,t) = P_\xi(\xi,t)P_\eta(\eta,t)$$
$$= \exp[-\frac{at}{2}[1-(\beta+\tau)t]\eta^2 - \frac{a\tau t}{2}\eta\xi - a r_1\tau t[1-\frac{rt}{4}]\eta^2 - \frac{at^2}{2}[1+(\frac{t}{r})^{-1}]\xi\eta - \frac{ar\tau t^3}{6}[1-\frac{3}{2}(rt)^{-1}]\xi^2] \ . \quad (2\text{-}43)$$

By using the inverse Fourier transform, the probability density $P(x,t)$ and $P(v,t)$ are, respectively, presented by

$$P(x,t) = [2\pi \frac{ar\tau t^3}{3}[1-\frac{3}{2}(rt)^{-1}]]^{-1/2} \exp[-\frac{x^2}{2\frac{ar\tau t^3}{3}[1-\frac{3}{2}(rt)^{-1}]}] \quad (2\text{-}44)$$

$$P(v,t) = [2\pi at[1-(\beta+\tau)t]]^{-1/2} \exp[-\frac{x^2}{2at[1-(\beta+\tau)t]}] \ . \quad (2\text{-}45)$$

The mean squared displacement and the mean squared velocity is, respectively, given by

$$<x^2(t)> = \frac{ar\tau t^3}{3}[1-\frac{3}{2}(rt)^{-1}], \quad <v^2(t)> = at[1-(\beta+\tau)t] \ . \quad (2\text{-}46)$$

The calculated process is complicated, but the moments of probability distribution function for $x$ and $v$ are surprisingly simple form, respectively, proportional to $t$ and $t^2$. The statistical values scales to $\sim t^3$ for the kurtosis. Our result for $t \ll \tau$ is different from that applied the random force [47].

**2-4 $P(x,t)$ and $P(v,t)$ with harmonic motion in $\tau = 0$ domain**

In $\tau = 0$ domain ($a(t) = 1, b(t) = 0$), we write approximate equation from Eq. (2-9) and (2-10) for $\xi$ and $\eta$

$$\frac{\partial}{\partial t} P_\xi(\xi,t) = -\beta\eta \frac{\partial}{\partial \xi} P_\xi(\xi,t) - \frac{1}{2}\alpha\eta^2 P_\xi(\xi,t) \tag{2-47}$$

$$\frac{\partial}{\partial t} P_\eta(\eta,t) = (\xi - r\eta)\frac{\partial}{\partial v}\eta P_\eta(\eta,t) - \frac{1}{2}\alpha\eta^2 P_\eta(\eta,t) \tag{2-48}$$

In steady state, we can calculate $P_\xi^{st}(\xi,t)$ and $P_\eta^{st}(\xi,t)$ as

$$P_\xi^{st}(\xi,t) = \exp[-\frac{\alpha}{2\beta\eta}\eta^2\xi - \frac{A}{\beta\eta}\xi], \quad P_\eta^{st}(\eta,t) = \exp[\frac{\alpha}{2\xi}\frac{\eta^3}{3} + \frac{\alpha r}{2\xi^2}\frac{\eta^4}{4} + \frac{A}{\xi}\eta + \frac{Ar}{\xi^2}\frac{\eta^3}{3}]. \tag{2-49}$$

We can find $P_\xi(\xi,t)$ and $P_\eta(\xi,t)$ as

$$P_\xi(\xi,t) = \Theta_\xi[t - \frac{\xi}{\beta\eta}]P^{st}(\xi,t), \quad P_\eta(\eta,t) = \Theta_\eta[t + \frac{\eta}{(\xi-r\eta)}]P_\eta^{st}(\eta,t). \tag{2-50}$$

Therefore, we calculate $P(\xi,\eta,t)$ from Eq. (44) and (45) as

$$P(\xi,\eta,t) = P_\xi(\xi,t)P_\eta(\eta,t)$$

$$= \exp[-\frac{\alpha rt^4}{8}[1+\frac{4}{3}[rt]^{-1}]\xi^2 - \frac{\alpha rt^3}{2}[1+[rt]^{-1}]\xi\eta - \frac{\alpha t}{2}[1-\frac{1}{4}[rt]^{-1}]\eta^2] \tag{2-51}$$

Using the inverse Fourier transform, $P(x,t)$ and $P(v,t)$ are, respectively, presented by

$$P(x,t) = [2\pi\frac{\alpha rt^4}{4}[1+\frac{4}{3}[rt]^{-1}]]^{-1/2} \exp[-\frac{x^2}{\alpha rt^4}[1+\frac{4}{3}[rt]^{-1}]^{-1}] \tag{2-52}$$

$$P(v,t) = [2\pi\alpha t[1-\frac{1}{4}[rt]^{-1}]]^{-1/2} \exp[-\frac{v^2}{2\alpha t}[1-\frac{1}{4}[rt]^{-1}]^{-1}] \tag{2-53}$$

with the mean squared deviations

$$<x^2(t)> = \frac{\alpha rt^4}{4}[1+\frac{4}{3}[rt]^{-1}], \quad <v^2(t)> = \alpha t[1-\frac{1}{4}[rt]^{-1}] \tag{2-54}$$

**2-5 Moment equation**

We study the moment equation for $\mu_{m,n}$ of distribution $P(x,v,t)$

$$\frac{d\mu_{m,n}}{dt} = rm\mu_{m-1,n+1} - rn\mu_{m,n} + \beta n\mu_{m+1,n-1} - mnac(t)\mu_{m-1,n-1} + n(n-1)ab(t)\mu_{m,n-2}. \tag{2-55}$$

Here $\mu_{m,n} = \int_{-\infty}^{+\infty} dx \int_{-\infty}^{+\infty} dv\, x^m v^n P(x,v,t)$.

The moment equation for $\mu_{m,n}$ of distribution $P(x,v,t)$ with viscous force and perturbative force $cx^3(t)$ as follows:

$$\frac{d\mu_{m,n}}{dt} = -r_1 m\mu_{m-1,n+1} + \alpha_2 m\mu_{m,n+1} + \frac{\alpha_2}{(1-r_2)}n\mu_{m+1,n-1} + \frac{\alpha}{(1-r_2)}mnb(t)\mu_{m-1,n-1} + \frac{\alpha}{(1-r_2)}n(n-1)a(t)\mu_{m,n-2}$$

(2-56)

To this end the kurtosis for $x$ and $v$ are, respectively, given by

$$K_x = <x^4>/3<x^2>^2 - 1, \quad K_v = <v^4>/3<v^2>^2 - 1. \quad (2\text{-}57)$$

We calculate the correlation coefficient,

$$\rho_{xv} = <(x-\mu_x)(v-\mu_v)>/\sigma_x \sigma_v \quad (2\text{-}58)$$

Here we assume that a passive particle is initially at $x = x_0$, $v = v_0$. The parameter $\mu_x$, $\mu_v$ denote the mean displacement and velocity of joint probability density, and $\sigma_x, \sigma_v$ the root-mean-squared displacement and velocity for a passive particle of joint probability density.

**Table 1**: Approximate values of the kurtosis, the correlation coefficient, and moment $\mu_{2,2}$ in three-time regions, for a passive particle with harmonic and viscous forces.

| Time domain | Variables | $K_x, K_v$ | $\rho_{xv}$ | $\mu_{2,2}$ |
|---|---|---|---|---|
| $t \ll \tau$ | $x$ | $\dfrac{\beta^2 \tau^4 x_0^4}{a^2} t^{-6} - \dfrac{\beta \tau^2 x_0^2}{a} t^{-3}$ | $\dfrac{\sqrt{\beta \tau^3} x_0 v_0}{a} t^{-5/2}$ | $\dfrac{a^2}{\beta \tau^3 (1+2rt)} t^5$ |
| $t \ll \tau$ | $v$ | $\dfrac{\tau^2 v_0^4}{a^2} t^{-4} - \dfrac{v_0^2}{a} t^{-2}$ | | |
| $t \gg \tau$ | $x$ | $\dfrac{x_0^4}{(ar\tau)^2} t^{-6} - \dfrac{x_0^2}{ar\tau} t^{-3}$ | $\dfrac{x_0 v_0}{\sqrt{r\tau} a} t^{-2}$ | $\dfrac{a^2 r\tau}{(1+2rt)} t^4$ |
| $t \gg \tau$ | $v$ | $\dfrac{v_0^4}{a^2} t^{-2} - \dfrac{v_0^2}{a} t^{-1}$ | | |
| $\tau = 0$ | $x$ | $\dfrac{x_0^4}{a^2 r^2} t^{-8} - \dfrac{x_0^2}{ar} t^{-4}$ | $\dfrac{x_0 v_0}{\sqrt{r} a} t^{-5/2}$ | $\dfrac{a^2 r}{(1+2rt)} t^5$ |
| $\tau = 0$ | $v$ | $\dfrac{v_0^4}{a^2} t^{-2} - \dfrac{v_0^2}{a} t^{-1}$ | | |

**Table 2**: Approximate values of the kurtosis, the correlation coefficient, and moment $\mu_{2,2}$ in three-time regions, for a passive particle with viscous force and perturbative force $cx^3(t)$.

| Time domain | Variable | $K_x, K_v$ | $\rho_{xv}$ | $\mu_{2,2}$ |
|---|---|---|---|---|
| $t \ll \tau$ | $x$ | $\dfrac{\tau^2 x_0^4}{a^2} t^{-8} - \dfrac{\tau x_0^2}{a} t^{-4}$ | $\dfrac{\tau x_0 v_0}{a} t^{-3}$ | $\dfrac{a^2}{(1+2rt)\tau^2} t^6$ |
| $t \ll \tau$ | $v$ | $\dfrac{\tau^2 v_0^4}{a^2} t^{-4} - \dfrac{\tau v_0^2}{a} t^{-2}$ | | |
| $t \gg \tau$ | $x$ | $\dfrac{\tau^2 x_0^4}{a^2} t^{-8} - \dfrac{\tau x_0^2}{a} t^{-4}$ | $\dfrac{\tau x_0 v_0}{a} t^{-3}$ | $\dfrac{a^2}{(1+2rt)\tau} t^5$ |
| $t \gg \tau$ | $v$ | $\dfrac{\tau^2 v_0^4}{a^2} t^{-4} - \dfrac{\tau v_0^2}{a} t^{-2}$ | | |
| $\tau = 0$ | $x$ | $\dfrac{x_0^4}{a^2 r^2} t^{-8} - \dfrac{x_0^2}{ar} t^{-4}$ | $\dfrac{x_0 v_0}{a} t^{-5/2}$ | $\dfrac{a^2}{(1+2rt)\tau} t^5$ |
| $\tau = 0$ | $v$ | $\dfrac{v_0^4}{a^2} t^{-2} - \dfrac{v_0^2}{a} t^{-1}$ | | |

# III. Fokker-Planck equation of a run-and-tumble particle

## 3-1 Fokker-Planck equation of displacement and velocity

A run-and-tumble particle of displacement and velocity is expressed in terms of

$$\frac{d}{dt}x(t) = v(t)\sigma(t), \quad \frac{d}{dt}v(t) = -rv(t) - \beta x(t) + g(t). \tag{3-1}$$

Here $-rv(t)$ denotes the consuming force acting on the particles, $r$ the viscous coefficient and $-\beta x(t)$ the harmonic force. A run-and-tumble particle acts on the motion that $\sigma(t)$ switches $\pm 1$ (+1 denotes a clockwise direction and $-1$ denotes a counter-clock direction) at rate $\gamma$. The value $g(t)$ is the force that activates the motion of the particle by the fluctuation of the particle. In Eq. (3-1), $g(t)$ denotes the fluctuation term of an exponentially correlated Gaussian force: $<g(t)> = 0$ and $<g(t)g(t')> = \frac{g_0^2}{2\tau}\exp(-\frac{|t-t'|}{\tau})$. Here $g_0^2 = 2rk_B T_{g_0}$, the parameter $g_0$ denotes the coupling strength, $\tau$ the correlation time, $k_B$ Boltzmann constant, $T$ temperature, and $r$ the frictional constant. Eq. (3-1) is reduced to

$$\frac{d}{dt}x_\pm(t) = \pm v(t)\gamma, \quad \frac{d}{dt}v_\pm(t) = -rv(t) - \beta x(t) + g(t) \tag{3-2}$$

The joint probability distribution function $P_\pm(x(t), v(t), t)$ for the displacement $x$ and the velocity $v$ is defined by $P_\pm(x(t), v(t), t) = <\delta(x - x_\pm(t))\delta(v - v_\pm(t))>$. We assume that the particle is initially at rest at $t = 0$. By taking time derivatives of the joint probability density, we have

$$\frac{\partial P_\pm(x,v,t)}{\partial t} = -\frac{\partial}{\partial x}<\frac{\partial x}{\partial t}\delta(x - x_\pm(t))\delta(v - v_\pm(t))> - \frac{\partial}{\partial v}<\frac{\partial v}{\partial t}\delta(x - x_\pm(t))\delta(v - v_\pm(t))>. \tag{3-3}$$

Inserting Eq. (3-2) and Eq. (3-3), we can write

$$\frac{\partial P_\pm(x,v,t)}{\partial t} = \mp\gamma\frac{\partial}{\partial x}<v\delta(x - x_\pm(t))\delta(v - v_\pm(t))> + <[rv(t) + \beta x(t) - g(t)]\delta(x - x_\pm(t))\delta(v - v_\pm(t))>. \tag{3-4}$$

By manipulating over integrals using Eq. (3-4) [47], the joint probability density $P(x(t), v(t), t)$ is derived as

$$\frac{\partial}{\partial t}P_\pm(x,v,t) = \mp\gamma v\frac{\partial}{\partial x}P_\pm(x,v,t) + r\frac{\partial}{\partial v}vP_\pm(x,v,t) + \beta x\frac{\partial}{\partial v}P_\pm(x,v,t)$$
$$-ac(t)\frac{\partial^2}{\partial x \partial v}P_\pm(x,v,t) + ab(t)\frac{\partial^2}{\partial v^2}P_\pm(x,v,t) \tag{3-5}$$

Here $a = g_0^2/2$, $c(t) = (t+\tau)\exp(-t/\tau) - \tau$ and $b(t) = 1 - \exp(-t/\tau)$.

## 3-2 $P_\pm(x,t)$ and $P_\pm(v,t)$ in short-time domain

In order to find the joint probability distribution function, we introduce the double Fourier transform

$$P_\pm(\xi, \eta, t) = \int_{-\infty}^{+\infty}dx\int_{-\infty}^{+\infty}dt\exp(-i\xi x_\pm - i\eta v_\pm)P_\pm(x,v,t). \tag{3-6}$$

By taking Fourier transform in Eq. (3-5), the Fourier transforms of joint probability distribution function $P_+(\xi,\eta,t)$ and $P_-(\xi,\eta,t)$ are, respectively, given by

$$\frac{\partial}{\partial t}P_+(\xi,\eta,t) = (\gamma\xi - r\eta)\frac{\partial}{\partial \eta}P_+(\xi,\eta,t) - \beta\eta\frac{\partial}{\partial \xi}P_+(\xi,\eta,t) + a[c(t)\xi\eta - b(t)\eta^2]P_+(\xi,\eta,t). \quad (3\text{-}7)$$

$$\frac{\partial}{\partial t}P_-(\xi,\eta,t) = -(\gamma\xi + r\eta)\frac{\partial}{\partial \eta}P_-(\xi,\eta,t) - \beta\eta\frac{\partial}{\partial \xi}P_-(\xi,\eta,t) + a[c(t)\xi\eta - b(t)\eta^2]P_-(\xi,\eta,t). \quad (3\text{-}8)$$

For $P_+(\xi,\eta,t)$ in steady state, introducing $\frac{\partial}{\partial t}P_+(\xi,\eta,t) = 0$ and $P_+(\xi,\eta,t) \to P_+^{st}(\xi,\eta,t)$, then we can calculate as

$$\gamma(\xi - \bar{r}\eta)\frac{\partial}{\partial \eta}P_+^{st}(\xi,\eta,t) - \beta\eta\frac{\partial}{\partial \xi}P_+^{st}(\xi,\eta,t) + a[c(t)\xi\eta - b(t)\eta^2]P_+^{st}(\xi,\eta,t) = 0. \quad (3\text{-}9)$$

Here the parameter $\bar{r} = r/\gamma$. Two equations in respect with obtaining the special solutions for $\xi$, $\eta$ by variable separation is as follows:

$$-\beta\eta\frac{\partial}{\partial \xi}P_{+\xi}^{st}(\xi,t) + \frac{1}{2}a[c(t)\xi\eta - b(t)\eta^2]P_{+\xi}^{st}(\xi,t) + AP_{+\xi}^{st}(\xi,t) = 0 \quad (3\text{-}10)$$

$$\gamma(\xi - \bar{r}\eta)\frac{\partial}{\partial \eta}P_{+\eta}^{st}(\eta,t) + \frac{1}{2}a[c(t)\xi\eta - b(t)\eta^2]P_{+\eta}^{st}(\eta,t) - AP_{+\eta}^{st}(\eta,t) = 0. \quad (3\text{-}11)$$

Here $A$ is the separation constant. From Eq. (3-10), Fourier transform of the probability distribution function for the velocity in the steady state is obtained as

$$P_{+\xi}^{st}(\xi,t) = \exp[\frac{a}{2\beta\eta}[\frac{c(t)}{2}\eta\xi^2 - b(t)\eta^2\xi]]. \quad (3\text{-}12)$$

In order to find the solutions of joint functions for $\xi$ from $P_{+\xi}(\xi,t) \equiv Q_{+\xi}(\xi,t)P_{+\xi}^{st}(\xi,t)$, we obtain the distribution functions in the short-time domain $t \ll \tau$ via the calculation including terms up to order $t^2/\tau^2$. That is,

$$P_{+\xi}(\xi,t) = Q_{+\xi}(\xi,t)\exp[\frac{a}{2\beta\eta}[\frac{c(t)}{2}\eta\xi^2 - b(t)\eta^2\xi]] \quad (3\text{-}13)$$

$$Q_{+\xi}(\xi,t) = R_{+\xi}(\xi,t)\exp[\frac{a}{2(\beta\eta)^2}[\frac{b'(t)}{2}\eta^2\xi^2 - \frac{c'(t)}{6}\eta\xi^3]] \quad (3\text{-}14)$$

$$R_{+\xi}(\xi,t) = S_{+\xi}(\xi,t)\exp[-\frac{a}{2(\beta\eta)^3}[\frac{b''(t)}{6}\eta^2\xi^3 - \frac{c''(t)}{24}\eta\xi^4]] \quad (3\text{-}15)$$

$$S_{+\xi}(\xi,t) = T_{+\xi}(\xi,t)\exp[\frac{a}{2(\beta\eta)^4}[\frac{b'''(t)}{24}\eta^2\xi^4 - \frac{c'''(t)}{120}\eta\xi^5]]. \quad (3\text{-}16)$$

Neglecting terms proportional to $1/\tau^3$ and taking the solutions as arbitrary functions of variable $t - (\xi/\beta\eta)$, the arbitrary function $T_{+\xi}(\xi,t)$ becomes $T_{+\xi}(\xi,t) = \Theta_\xi[t - (\xi/\beta\eta)]$. We consequently find that

$$P_{+\xi}(\xi,t) = \Theta_\xi[t - (\xi/\beta\eta)]S_{+\xi}^{st}(\xi,t)R_{+\xi}^{st}(\xi,t)Q_{+\xi}^{st}(\xi,t)P_{+\xi}^{st}(\xi,t). \quad (3\text{-}17)$$

By expanding their derivatives to second order in powers of $t/\tau$, we obtain the expression for $P_{+\xi}(\xi,t)$ after some cancellations. That is,

$$P_{+\xi}(\xi,t) = \exp[-\frac{at^3}{12\beta\tau^2}[1 - 3[\frac{t}{\tau}]^{-1}]\xi^2 - \frac{at^3}{4\tau}[1 - \frac{1}{6}[\frac{t}{\tau}]]\eta\xi - \frac{a\beta t^3}{4}[1 - \frac{1}{3}[\frac{t}{\tau}]]\eta^2]. \quad (3\text{-}18)$$

Here,

$$\Theta_\xi[u] = \exp[-\frac{a\beta}{120\tau^2}\eta^2 u^5 + \frac{a\beta}{48\tau}\eta^2 u^4 + \frac{a\beta t}{12\tau}\eta^2 u^3 - \frac{a}{4\tau}\eta^2 u^2 - \frac{a\beta t}{4}\eta^2 u^2]. \quad (3\text{-}19)$$

By using the inverse Fourier transform, the probability density $P(x_+,t)$ is presented by

$$P(x_+,t) = \frac{1}{2\pi}\int_{-\infty}^{+\infty} d\xi \exp(-i\xi x_+) P_{+\xi}(\xi,t)$$

$$= [\pi \frac{at^3}{3\beta\tau^2}[1-3[\frac{t}{\tau}]^{-1}]]^{-1/2} \exp[-\frac{3\beta\tau^2 x^2}{at^3}[1-3[\frac{t}{\tau}]^{-1}]]. \quad (3\text{-}20)$$

The mean squared displacement for $P(x_+,t)$ is given by

$$<x_+^2(t)> = \frac{at^3}{6\beta\tau^2}[1-3[\frac{t}{\tau}]^{-1}]. \quad (3\text{-}21)$$

For obtaining the special solutions for $\eta$ in the short-time domain $t \ll \tau$, we use the variable-separated equation. In the steady state, we assume that $\frac{1}{\gamma(\xi-\overline{r}\eta)} \cong \frac{1}{\gamma\xi}(1+\frac{\overline{r}\eta}{\xi})$. We calculate Eq. (3-11) in the steady state as

$$P_{+\eta}^{st}(\eta,t) = \exp[-\frac{a}{2\gamma\xi}[-\frac{b(t)}{3}\eta^3 + \frac{c(t)}{2}\xi\eta^2] + \frac{a\overline{r}}{2\gamma\xi^2}[\frac{b(t)}{4}\eta^4 - \frac{c(t)}{3}\xi\eta^3]]. \quad (3\text{-}22)$$

From similar method, we obtain from $Q_\eta(\eta,t)P_\eta^{st}(\eta,t)$ that

$$P_{+\eta}(\eta,t) = Q_{+\eta}(\eta,t)P_{+\eta}^{st}(\eta,t) \quad (3\text{-}23)$$

$$Q_{+\eta}(\eta,t) = R_{+\eta}(\eta,t)\exp[\frac{a}{2\gamma^2\xi^2}[\frac{b'(t)}{12}\eta^4 - \frac{c'(t)}{6}\xi\eta^3] + \frac{a\overline{r}}{2\gamma^2\xi^3}[\frac{b'(t)}{20}\eta^5 - \frac{c'(t)}{12}\xi\eta^4]] \quad (3\text{-}24)$$

$$R_{+\eta}(\eta,t) = S_{+\eta}(\eta,t)\exp[\frac{a}{2\gamma^3\xi^3}[\frac{b''(t)}{60}\eta^5 - \frac{c''(t)}{24}\xi\eta^4] + \frac{a\overline{r}}{2\gamma^3\xi^4}[\frac{b''(t)}{120}\eta^6 - \frac{c''(t)}{60}\xi\eta^5]] \quad (3\text{-}25)$$

$$S_{+\eta}(\eta,t) = T_{+\eta}(\eta,t)\exp[-\frac{a}{\gamma^4\xi^3}\frac{c'''(t)}{240}\eta^5 - \frac{a\overline{r}}{\gamma^4\xi^4}\frac{c'''(t)}{720}\eta^6]. \quad (3\text{-}26)$$

Here, $b'(t) = db(t)/dt$ and $c'(t) = dc(t)/dt$. Neglecting terms proportional to $1/\tau^3$, $T_{+\eta}(\eta,t)$ is given by

$$T_{+\eta}(\eta,t) = \gamma(\xi-\overline{r}\eta)\frac{\partial}{\partial\eta}T_\eta(\eta,t). \quad (3\text{-}27)$$

Taking the solutions as arbitrary functions of variable $t+\eta/\gamma(\xi-\overline{r}\eta)$, the arbitrary function become $T_{+\eta}(\eta,t) = \Theta_\eta[t+[\eta/\gamma(\xi-\overline{r}\eta)]]$. Consequently, expanding their derivatives to second order in powers of $t/\tau$, we respectively obtain the expression for $P_{+\eta}(\eta,t)$ after some cancellations. That is,

$$P_{+\eta}(\eta,t) = \Theta_\eta[t+\eta/\gamma(\xi-\overline{r}\eta)]S_{+\eta}^{st}(\eta,t)R_{+\eta}^{st}(\eta,t)Q_{+\eta}^{st}(\eta,t)P_{+\eta}^{st}(\eta,t)$$

$$= \exp[-\frac{3a\overline{r}^{-2}t^4}{\gamma\tau}[1+\frac{\gamma}{6rt}]\eta^2 - \frac{a\overline{r}t^4}{4\gamma^2\tau}[1+\frac{2\gamma}{rt}]\xi\eta - \frac{at^4}{6\gamma\tau}[1-\frac{3\tau}{2t}]\xi^2]. \quad (3\text{-}28)$$

Using the inverse Fourier transform, $P(v_+,t)$ is presented by

$$P(v_+,t) = \frac{1}{2\pi}\int_{-\infty}^{+\infty} d\eta \exp(-i\eta v_+)P_{+\eta}(\eta,t) = [2\pi \frac{6a\bar{r}^{-2}t^4}{\gamma\tau}[1+\frac{\gamma}{6rt}]]^{-1/2}\exp[-\frac{\gamma\tau v_+^2}{12a\bar{r}^{-2}t^4}[1+\frac{\gamma}{6rt}]^{-1}]. \quad (3\text{-}29)$$

The mean squared velocity for the probability density $P(v,t)$ is given by

$$<v_+^2(t)> = \frac{6a\bar{r}^{-2}t^4}{\gamma\tau}[1+\frac{\gamma}{6rt}]. \quad (3\text{-}30)$$

For $P_-(\xi,\eta,t)$ in steady state, introducing $\frac{\partial}{\partial t}P_-(\xi,\eta,t)=0$ and $P_-(\xi,\eta,t) \to P_-^{st}(\xi,\eta,t)$, then we can calculate as

$$-(\xi+\bar{r}\eta)\frac{\partial}{\partial \eta}P_-^{st}(\xi,\eta,t) - \beta\eta\frac{\partial}{\partial \xi}P_-^{st}(\xi,\eta,t) + a[c(t)\xi\eta - b(t)\eta^2]P_-^{st}(\xi,\eta,t) = 0. \quad (3\text{-}31)$$

Two equations in respect with obtaining the special solutions for $\xi$, $\eta$ by variable separation is as follows:

$$-\beta\eta\frac{\partial}{\partial \xi}P_{-\xi}^{st}(\xi,t) + \frac{1}{2}a[c(t)\xi\eta - b(t)\eta^2]P_{-\xi}^{st}(\xi,t) + BP_{-\xi}^{st}(\xi,t) = 0 \quad (3\text{-}32)$$

$$-(\gamma\xi+\bar{r}\eta)\frac{\partial}{\partial \eta}P_{-\eta}^{st}(\eta,t) + \frac{1}{2}a[c(t)\xi\eta - b(t)\eta^2]P_{-\eta}^{st}(\eta,t) - BP_{-\eta}^{st}(\eta,t) = 0. \quad (3\text{-}33)$$

When we consider to find the probability density $P_-(x,t)$ from Eq. (3-32), the method finding the solution of Eq. (3-32) has the same as the method of Eq. (3-20). Eq. (3-20) (probability density for $x_-(t)$) becomes a solution of Eq. (3.32), i.e. $P(x_+,t) = P(x_-,t)$.

In order to obtain the special solution $P_-(\eta,t)$ for $\eta$ in the short-time domain $t \ll \tau$, we use the variable-separated equation, Eq. (3-33). Assuming that $\frac{1}{\gamma(\xi+\bar{r}\eta)} \cong \frac{1}{\gamma\xi}(1-\frac{\bar{r}\eta}{\xi})$, we calculate $P_{-\eta}^{st}(\eta,t)$ as

$$P_{-\eta}^{st}(\eta,t) = \exp[\frac{a}{2\gamma\xi}[-\frac{b(t)}{3}\eta^3 + \frac{c(t)}{2}\xi\eta^2] + \frac{a\bar{r}}{2\gamma\xi^2}[\frac{b(t)}{4}\eta^4 - \frac{c(t)}{3}\xi\eta^3]]. \quad (3\text{-}34)$$

From similar method, we obtain from $Q_\eta(\eta,t)P_\eta^{st}(\eta,t)$ that

$$P_{-\eta}(\eta,t) = Q_{-\eta}(\eta,t)P_{-\eta}^{st}(\eta,t) \quad (3\text{-}35)$$

$$Q_{-\eta}(\eta,t) = R_{-\eta}(\eta,t)\exp[\frac{a}{2\gamma^2\xi^2}[\frac{b'(t)}{12}\eta^4 - \frac{c'(t)}{6}\xi\eta^3] - \frac{a\bar{r}}{2\gamma^2\xi^3}[\frac{b'(t)}{4}\eta^5 - \frac{c'(t)}{3}\xi\eta^4]] \quad (3\text{-}36)$$

$$R_{-\eta}(\eta,t) = S_{-\eta}(\eta,t)\exp[-\frac{a}{2\gamma^3\xi^3}[\frac{b''(t)}{60}\eta^5 - \frac{c''(t)}{24}\xi\eta^4] + \frac{a\bar{r}}{2\gamma^3\xi^4}[\frac{b''(t)}{120}\eta^6 - \frac{c''(t)}{60}\xi\eta^6]] \quad (3\text{-}37)$$

$$S_{-\eta}(\eta,t) = T_{-\eta}(\eta,t)\exp[-\frac{a}{\gamma^4\xi^3}\frac{c'''(t)}{240}\eta^5 + \frac{a\bar{r}}{\gamma^4\xi^4}\frac{c'''(t)}{720}\eta^6]. \quad (3\text{-}38)$$

Neglecting terms proportional to $1/\tau^3$, $T_{-\eta}(\eta,t)$ is given by

$$T_{-\eta}(\eta,t) = -\gamma[\xi+\bar{r}\eta]\frac{\partial}{\partial \eta}T_{-\eta}(\eta,t). \quad (3\text{-}39)$$

Taking the solutions as arbitrary functions of variable $t-[\eta/\gamma(\xi+\bar{r}\eta)]$, the arbitrary function become $T_{-\eta}(\eta,t) = \Theta_\eta[t-[\eta/\gamma(\xi+\bar{r}\eta)]]$. Consequently, expanding their derivatives to second order in powers of $t/\tau$, we get the expression for $P_{-\eta}(\eta,t)$ after some cancellations. That is,

$$P_{-\eta}(\eta,t) = \Theta_{\eta}[t-\eta/\gamma(\xi-\bar{r}\eta)]S^{st}_{-\eta}(\eta,t)R^{st}_{-\eta}(\eta,t)Q^{st}_{-\eta}(\eta,t)P^{st}_{-\eta}(\eta,t)$$

$$= \exp[-\frac{3a\bar{r}^{-2}t^4}{\gamma\tau}[1-\frac{1}{3rt}]\eta^2 - \frac{at^4}{12\gamma\tau}[1-\frac{3\tau}{t}]\xi\eta - \frac{a\bar{r}t^4}{4\tau}[1-\frac{2\tau}{t}]\xi^2]. \quad (3\text{-}40)$$

Using the inverse Fourier transform, the probability density $P_-(\eta,t)$ and $P_-(v,t)$ are, respectively, calculated as

$$P_-(\eta,t) = \exp[-\frac{at^3}{4}[1-\frac{1}{3}[\frac{t}{\tau}]^{-1}]\xi^2 - \frac{11}{24}\frac{a\bar{r}t^4}{\tau}[1+\frac{14}{11}\frac{1}{rt}]\xi\eta - \frac{3}{8}\frac{at^2}{\tau}[(1+\frac{4}{3}r)+\frac{20}{9}rt]\eta^2] \quad (3\text{-}41)$$

$$P(v_-,t) = [2\pi\frac{6a\bar{r}^{-2}t^4}{\gamma\tau}[1-\frac{1}{3rt}]]^{-1/2}\exp[-\frac{\gamma\tau v_-^2}{12a\bar{r}^{-2}t^4}[1-\frac{1}{3rt}]^{-1}]. \quad (3\text{-}42)$$

The mean squared displacement and velocity for $P_-(x,t)$ and $P_-(v,t)$ is, respectively, given by

$$<x_-^2(t)> = \frac{at^3}{3\beta\tau^2}[1-3[\frac{t}{\tau}]^{-1}], \quad <v_-^2(t)> = \frac{6a\bar{r}^{-2}t^4}{\gamma\tau}[1-\frac{1}{3rt}]. \quad (3\text{-}43)$$

### 3-3 $P(x,t)$ and $P(v,t)$ in long-time domain

In long-time domain, we write approximate equation for $\pm\xi$

$$\frac{\partial}{\partial t}P_{\pm\xi}(\xi,t) \cong \frac{a}{2}[c(t)\xi\eta - b(t)\eta^2]P_{\pm\xi}(\xi,t). \quad (3\text{-}44)$$

We can see that the method to find the solution of $P_{+\xi}(\xi,t)$ is the same as that to find the solution of $P_{+\xi}(\xi,t)$. In steady state, introducing $\frac{\partial}{\partial t}P_{\pm\xi}(\xi,t) = 0$ for $x$ and $P_{\pm\xi}(\xi,t) \to P^{st}_{\pm\xi}(\xi,t)$, then we can calculate $P^{st}_{\pm\xi}(\xi,t)$ as

$$P_{\pm\xi}(\xi,t) = \exp[\frac{a}{2}\int[c(t)\xi\eta - b(t)\eta^2]\,dt]. \quad (3\text{-}45)$$

So to speak, as $\int b(t)dt = t-\tau$, $b(t)=1$ and $\int c(t)dt = -\tau t$, $c(t)=-\tau$ in long-time domain, we can find $Q^{st}_{\pm\xi}(\xi,t)$ for $\xi$ from $P_{\pm\xi}(\xi,t) \equiv Q_{\pm\xi}(\xi,t)P^{st}_{\pm\xi}(\xi,t)$, that is,

$$Q^{st}_{\pm\xi}(\xi,t) = \exp[\frac{a}{2}\int[b(t)\eta^2 - c(t)\xi\eta]\,dt]. \quad (3\text{-}46)$$

Hence, we have $P^{st}_{\pm\xi}(\xi,t)$ and $Q_{\pm\xi}(\xi,t)$ for long-time domain $t \gg \tau$ as

$$P^{st}_{\pm\xi}(\xi,t) = \exp[\frac{a}{2\beta\eta}[\frac{c(t)}{2}\eta\xi^2 - b(t)\eta^2\xi]] \quad (3\text{-}47)$$

$$Q_{\pm\xi}(\xi,t) = R_{\pm\xi}(\xi,t)Q^{st}_{\pm\xi}(\xi,t) = R_{\pm\xi}(\xi,t)\exp[\frac{a}{2}\int[b(t)\eta^2 - c(t)\eta\xi]\,dt]. \quad (3\text{-}48)$$

Taking the solutions as arbitrary functions of variable $t-\xi/\beta\eta$, the arbitrary function $R_{\pm\xi}(\xi,t)$ becomes $R_{\pm\xi}(\xi,t) = \Theta_{\xi}[t-[\xi/\beta\eta]]$. Consequently, as expanding their derivatives to second order in powers of $t/\tau$, we obtain the expression for $P_{\pm}(\xi,t)$ after some cancellations.

That is,
$$P_{\pm}(\xi,t) = R_{\pm\xi}(\xi,t)Q_{\pm\xi}^{st}(\xi,t)P_{\pm\xi}^{st}(\xi,t) = \Theta_{\xi}[t-[\xi/\beta\eta]])Q_{\xi}^{st}(\xi,t)P_{\xi}^{st}(\xi,t). \quad (3\text{-}49)$$
Here,
$$\Theta_{\xi}(u) = \exp[\frac{a}{2}\beta\tau t\eta^2 u - \frac{a}{2}(t-\tau)\eta^2 + \frac{a\beta\tau}{4}v^2 u^2 - \frac{a}{2}\eta^2 u]. \quad (3\text{-}50)$$

In long-time domain, we write approximate equation for $\eta$
$$\frac{\partial}{\partial t}P_{\pm\eta}(\eta,t) \cong \frac{a}{2}[c(t)\xi\eta - b(t)\eta^2]P_{\pm\eta}(\eta,t). \quad (3\text{-}51)$$

We can simply calculate $P_{\pm\eta}(\eta,t)$ as
$$P_{\pm\eta}(\eta,t) = \exp[\frac{a}{2}\int[c(t)\xi\eta - b(t)\eta^2]\,dt. \quad (3\text{-}52)$$

We find $Q_{\pm\eta}^{st}(\eta,t)$ for $\eta$ from $P_{\pm\eta}(\eta,t) = Q_{\pm\eta}(\eta,t)P_{\pm\eta}^{st}(\eta,t)$, That is,
$$Q_{\pm\eta}^{st}(\eta,t) = \exp[\frac{a}{2}\int[b(t)\eta^2 - c(t)\xi\eta]\,dt. \quad (3\text{-}53)$$

In steady state for long-time domain, $P_{+\eta}^{st}(\eta,t)$ and $P_{-\eta}^{st}(\eta,t)$ are, respectively, calculated as
$$P_{+\eta}^{st}(\eta,t) = \exp[-\frac{a}{2\gamma\xi}[-\frac{b(t)}{3}\eta^3 + \frac{c(t)}{2}\xi\eta^2] + \frac{a\bar{r}}{2\gamma\xi^2}[\frac{b(t)}{4}\eta^4 - \frac{c(t)}{3}\xi\eta^3]] \quad (3\text{-}54)$$

$$P_{-\eta}^{st}(\eta,t) = \exp[\frac{a}{2\gamma\xi}[-\frac{b(t)}{3}\eta^3 + \frac{c(t)}{2}\xi\eta^2] + \frac{a\bar{r}}{2\gamma\xi^2}[\frac{b(t)}{4}\eta^4 - \frac{c(t)}{3}\xi\eta^3]]. \quad (3\text{-}55)$$

Taking the solutions as arbitrary functions of variable $t+\eta/\gamma[\xi-\bar{r}\eta]$, $t-\eta/\gamma[\xi+\bar{r}\eta]$, the arbitrary function $R_{\eta}(\eta,t)$ becomes $R_{+\eta}(\eta,t) = \Theta_{\eta}[t+\eta/\gamma[\xi-\bar{r}\eta]]$, $R_{-\eta}(\eta,t) = \Theta_{\eta}[t-\eta/\gamma[\xi+\bar{r}\eta]]$
Consequently, as expanding their derivatives to second order in powers of $t/\tau$, we obtain the expressions for $P_{+\eta}(\eta,t)$, $P_{-\eta}(\eta,t)$ after some cancellations, respectively. That is,
$$P_{+\eta}(\eta,t) = R_{+\eta}(\eta,t)Q_{+\eta}^{st}(\eta,t)P_{+\eta}^{st}(\eta,t) = \Theta_{\xi}[t+\eta/\gamma[\xi-\bar{r}\eta]]Q_{+\eta}^{st}(\eta,t)P_{+\eta}^{st}(\eta,t) \quad (3\text{-}56)$$
$$P_{-}(\eta,t) = R_{-\eta}(\eta,t)Q_{-\eta}^{st}(\eta,t)P_{-\eta}^{st}(\eta,t) = \Theta_{\xi}[t-\eta/\gamma[\xi+\bar{r}\eta]]Q_{-\eta}^{st}(\eta,t)P_{-\eta}^{st}(\eta,t). \quad (3\text{-}57)$$

The probability density $P_{+}(\xi,\eta,t)$ and $P_{-}(\xi,\eta,t)$ are, respectively, calculated as
$$P_{+}(\xi,\eta,t) = P_{+}(\xi,t)P_{+}(\eta,t)$$
$$= \exp[-\frac{a\bar{r}t^4}{8\gamma\tau}[1+\frac{4\tau}{3\bar{r}t}]\xi^2 - [\frac{at^2}{2} + \frac{a\bar{r}t^3}{6\gamma\tau}]\xi\eta - \frac{at}{2}[1-\beta t]\eta^2] \quad (3\text{-}58)$$

$$P_{-}(\xi,\eta,t) = P_{-}(\xi,t)P_{-}(\eta,t)$$
$$= \exp[-\frac{a\bar{r}t^4}{8\gamma\tau}[1-\frac{4\tau}{3\bar{r}t}]\xi^2 - [\frac{a\beta t^2}{2} + \frac{a\bar{r}t^3}{6\gamma\tau}]\xi\eta - \frac{at}{\tau}[1-2\beta\tau t]\eta^2] \quad (3\text{-}59)$$

By using the inverse Fourier transform, the probability density $P(x_{\pm},t)$ and $P(v_{\pm},t)$ are, respectively, presented by
$$P(x_{+},t) = [\pi\frac{a\bar{r}t^4}{2\gamma\tau}[1+\frac{4\tau}{3\bar{r}t}]]^{-1/2}\exp[-\frac{2\gamma\tau x_{+}^2}{a\bar{r}t^4}[1+\frac{4\tau}{3\bar{r}t}]^{-1}] \quad (3\text{-}60)$$

$$P(v_{+},t) = [2\pi at[1-\beta t]]^{-1/2}\exp[-\frac{v_{+}^2}{2at}[1-\beta t]^{-1}] \quad (3\text{-}61)$$

$$P(x_-,t) = [\pi \frac{\overline{ar}t^4}{2\gamma\tau}[1-\frac{4\tau}{3\overline{r}t}]]^{-1/2} \exp[-\frac{2\gamma\tau x_-^2}{\overline{ar}t^4}[1-\frac{4\tau}{3\overline{r}t}]^{-1}] \tag{3-62}$$

$$P(v_-,t) = [4\pi \frac{at}{\tau}[1-2\beta\tau t]]^{-1/2} \exp[-\frac{\tau v_-^2}{4at}[1-2\beta\tau t]^{-1}]. \tag{3-63}$$

The mean squared displacement and the mean squared velocity for $P_\pm(x,t)$ and $P_\pm(v,t)$ is, respectively, given by

$$<x_+^2(t)> = \frac{\overline{ar}t^4}{4\gamma\tau}[1+\frac{4\tau}{3\overline{r}t}], \quad <v_+^2(t)> = at[1-\beta t] \tag{3-64}$$

$$<x_-^2(t)> = \frac{\overline{ar}t^4}{4\gamma\tau}[1-\frac{4\tau}{3\overline{r}t}], \quad <v_-^2(t)> = \frac{2at}{\tau}[1-2\beta\tau t]. \tag{3-65}$$

### 3-4 $P_\pm(x,t)$ and $P_\pm(v,t)$ in $\tau=0$ domain

In $\tau=0$ domain ($a(t)=1, b(t)=0$), we write approximate equation for $\xi$ and $\eta$ as

$$\frac{\partial}{\partial t}P_\pm(\xi,t) = -\beta\eta\frac{\partial}{\partial \xi}P_\pm(\xi,t) - \frac{1}{2}a\eta^2 P_\pm(\xi,t) \tag{3-66}$$

$$\frac{\partial}{\partial t}P_+(\eta,t) = (\xi - r\eta)\frac{\partial}{\partial v}\eta P_+(\eta,t) - \frac{1}{2}\alpha\eta^2 P_+(\eta,t) + AP_+(\eta,t) \tag{3-67}$$

$$\frac{\partial}{\partial t}P_-(\eta,t) = -(\xi + r\eta)\frac{\partial}{\partial v}\eta P_-(\eta,t) - \frac{1}{2}\alpha\eta^2 P_-(\eta,t) - AP_-(\eta,t) \tag{3-68}$$

In steady state, we can calculate $P_\pm^{st}(\xi,t)$ and $P_\pm^{st}(\eta,t)$ as

$$P_\pm^{st}(\xi,t) = \exp[-\frac{\alpha}{2\beta\eta}\eta^2\xi + \frac{A}{\beta\eta}\xi] \tag{3-69}$$

$$P_+^{st}(\eta,t) = \exp[\frac{\alpha}{2\xi}\frac{\eta^3}{3} + \frac{\alpha r}{2\xi^2}\frac{\eta^4}{4} - \frac{A}{\xi}\eta + \frac{Ar}{\xi^2}\frac{\eta^3}{3}] \tag{3-70}$$

$$P_-^{st}(\eta,t) = \exp[-\frac{\alpha}{2\xi}\frac{\eta^3}{3} + \frac{\alpha r}{2\xi^2}\frac{\eta^4}{4} - \frac{A}{\xi}\eta + \frac{Ar}{\xi^2}\frac{\eta^3}{3}]. \tag{3-71}$$

We can find $P_\pm^{st}(\xi,t)$ and $P_\pm^{st}(\eta,t)$ as

$$P_\pm(\xi,t) = \Theta_\xi[t - \frac{\xi}{\beta\eta}]P_\pm^{st}(\xi,t) \tag{3-72}$$

$$P_+(\eta,t) = \Theta_\eta[t + \frac{\eta}{\gamma(\xi - r\eta)}]P_+^{st}(\eta,t) \tag{3-73}$$

$$P_-(\eta,t) = \Theta_\eta[t - \frac{\eta}{\gamma(\xi + r\eta)}]P_-^{st}(\eta,t). \tag{3-74}$$

The probability density $P_+(\xi,\eta,t)$ and $P_-(\xi,\eta,t)$ are, respectively, calculated as

$$P_+(\xi,\eta,t) = P_+(\xi,t)P_+(\eta,t)$$

$$= \exp[-\frac{\overline{ar}t^4}{8\gamma}[1+\frac{4}{3\overline{r}t}]\xi^2 - \frac{at^2}{2\gamma}\xi\eta - at[1+\frac{rt}{4}]\eta^2] \tag{3-75}$$

$$P_-(\xi,v,t) = P_-(\xi,t) + P_-(v,t)$$

$$= \exp[-\frac{\overline{art}^4}{8\gamma}[1-\frac{4}{3rt}]\xi^2 - \frac{art^3}{\gamma}[1-\frac{1}{2rt}]\xi\eta - 2at\eta^2]. \qquad (3\text{-}76)$$

By using the inverse Fourier transform, $P_{\pm}(x,t)$ and $P_{\pm}(v,t)$ are, respectively, presented by

$$P(x_+,t) = [\pi\frac{\overline{art}^4}{2\gamma}[1+\frac{4}{3rt}]]^{-1/2} \exp[-\frac{2\gamma x_+^2}{\alpha rt^4}[1+\frac{4}{3rt}]^{-1}] \qquad (3\text{-}77)$$

$$P(v_+,t) = [4\pi at[1+\frac{rt}{4}]]^{-1/2} \exp[-\frac{v_+^2}{8at^2}[1+\frac{rt}{4}]^{-1}] \qquad (3\text{-}78)$$

$$P(x_-,t) = [\pi\frac{\overline{art}^4}{2\gamma}[1-\frac{4}{3rt}]]^{-1/2} \exp[-\frac{2x_-^2}{\alpha rt^4}[1-\frac{4}{3rt}]^{-1}] \qquad (3\text{-}79)$$

$$P(v_-,t) = [8\pi at[1-\frac{8}{3}r^2t]]^{-1/2} \exp[-\frac{v_-^2}{8at^2}[1-\frac{8}{3}r^2t]^{-1}]. \qquad (3\text{-}80)$$

The mean squared displacement and the mean squared velocity for $P_{\pm}(x,t)$ and $P_{\pm}(v,t)$ are, respectively, given by

$$<x_+^2(t)> = \frac{\overline{art}^4}{4\gamma}[1+\frac{4}{3rt}], \quad <v_+^2(t)> = 2at[1+\frac{rt}{4}] \qquad (3\text{-}81)$$

$$<x_-^2(t)> = \frac{\overline{art}^4}{2\gamma}[1-\frac{4}{3rt}], \quad <v_+^2(t)> = 4at[1-\frac{8}{3}r^2t]. \qquad (3\text{-}82)$$

### 3-5 Moment equation

**Table 3**: Approximate values of the kurtosis, the correlation coefficient, and moment $\mu_{2,2}^{\pm}$ in three-time regions, for a run-and-tumble particle with harmonic and viscous forces.

| time | variable | $K_x^{\pm}, K_v^{\pm}$ | $\rho_{x,v}^{\pm}$ | $\mu_{2,2}^{\pm}$ |
|---|---|---|---|---|
| $t \ll \tau$ | $x$ | $K_x^+ \cong \frac{\beta^2\tau^4 x_0^4}{a^2}t^{-6}$; $K_x^- \cong \frac{\beta^2\tau^2 x_0^4}{a^2}t^{-6}$ | $\rho_{xv}^+ = \frac{(\beta\gamma\tau^3)^{1/2}x_0v_0}{ar}t^{-5/2}$ | $\mu_{2,2}^{\pm} \cong \frac{a^2}{(1+2rt)\beta\tau^2}t^5$ |
| $t \ll \tau$ | $v$ | $K_v^+ \cong \frac{\gamma^2\tau^2 v_0^4}{a^2 r}t^{-4}$; $K_v^- \cong \frac{\tau^2 v_0^4}{a^2}t^{-4}$ | $\rho_{xv}^- = \frac{\beta^{1/2}\tau x_0v_0}{a}t^{-5/2}$ | $\mu_{2,2}^{\pm} \cong \frac{a^2}{(1+2rt)\gamma\tau}t^6$ |
| $t \gg \tau$ | $x$ | $K_x^+ \cong \frac{\gamma^2\tau^2 x_0^4}{a^2 r^{-2}}t^{-8}$; $K_x^- \cong \frac{\gamma^2\tau^2 x_0^4}{a^2 r^{-2}}t^{-8}$ | $\rho_{xv}^+ = \frac{(\gamma\tau)^{1/2} x_0 v_0}{ar^{-1/2}}t^{-5/2}$ | $\mu_{2,2}^{\pm} \cong \frac{a^2 r}{(1+2rt)\gamma^2\tau}t^6$ |
| $t \gg \tau$ | $v$ | $K_v^+ \cong \frac{v_0^4}{a^2}t^{-2}$; $K_v^- \cong \frac{v_0^4}{a^2}t^{-2}$ | $\rho_{xv}^- = \frac{(\gamma\tau)^{1/2} x_0 v_0}{ar^{-1/2}}t^{-5/2}$ | $\mu_{2,2}^+ \cong \frac{a^2}{(1+2rt)}t^3$; $\mu_{2,2}^- \cong \frac{a^2}{(1+2rt)\tau}t^3$ |
| $\tau = 0$ | $x$ | $K_x^+ \cong \frac{\gamma^2 x_0^4}{a^2 r^{-2}}t^{-8}$; $K_x^- \cong \frac{\gamma^2 x_0^4}{a^2 r^{-2}}t^{-8}$ | $\rho_{xv}^+ = \frac{\gamma^{1/2} x_0 v_0}{ar^{-1/2}}t^{-5/2}$ | $\mu_{2,2}^{\pm} \cong \frac{a^2 r}{(1+2rt)\gamma^2}t^6$ |
| $\tau = 0$ | $v$ | $K_v^+ \cong \frac{v_0^4}{a^2}t^{-2}$; $K_v^- \cong \frac{v_0^4}{a^2}t^{-2}$ | $\rho_{xv}^- = \frac{\gamma^{1/2} x_0 v_0}{ar^{-1/2}}t^{-5/2}$ | $\mu_{2,2}^{\pm} \cong \frac{a^2}{(1+2rt)}t^3$ |

To this end we study the moment equation for $\mu_{m,n}^{\pm}$ of distribution $P_{\pm}(\xi,\eta,t)$

$$\frac{d\mu_{m,n}^{\pm}}{dt} = \pm\gamma\mu_{m-1,n+1}^{\pm} + m\mu_{m,n}^{\pm} - rn\mu_{m,n}^{\pm} + \beta n\mu_{m+1,n-1}^{\pm} - mnac(t)\mu_{m-1,n-1}^{\pm} + n(n-1)ab(t)\mu_{m,n-2}^{\pm}. \qquad (3\text{-}83)$$

Here $\mu_{m,n}^{\pm} = \int_{-\infty}^{+\infty}dx\int_{-\infty}^{+\infty}dv\,x^m v^n P_{\pm}(\xi,\eta,t)$. We can simply derive Eq. (3-83) using Eq. (3.5).

The kurtosis for $x$ and $v$ are, respectively, given by
$$K_x^\pm = <x_\pm^4(t)>/3<x_\pm^2(t)>^2 - 1, \quad K_v^\pm = <v_\pm^4(t)>/3<v_\pm^2(t)>^2 - 1. \tag{3-84}$$
We also calculate the correlation coefficient as
$$\rho_{x,v}^\pm = <(x_\pm(t)-\mu_{\pm x})(v_\pm(t)-\mu_{\pm v})>/\sigma_{\pm x}\sigma_{\pm v}. \tag{3-85}$$
Here we assume that for a run-and-tumble particle is initially at $x_\pm(t)=x_0$, $v_\pm(t)=v_0$. The parameter $\mu_{\pm x}$, $\mu_{\pm v}$ denote the mean displacement and velocity of joint probability density, and $\sigma_{\pm x}$, $\sigma_{\pm v}$ the root-mean-squared displacement and velocity.

## IV. Summary

In summary, we have derived the Fokker-Planck equation for a passive particle with harmonic, viscous, and perturbative forces. After deriving the Fokker-Planck equation for a run-and-tumble particle, we have approximately solved the solution for the joint distribution density subject to an exponential correlated Gaussian force in three kinds of time limit domains, that is, $t \ll \tau$, $t \gg \tau$ and for $\tau = 0$.

Mean squared velocity (displacement) for a passive particle having harmonic and viscous forces behaviors in the form of super-diffusion scales to $\sim t^2$ ($\sim t^4$) in $t \ll \tau$ ($\tau = 0$), consistent with that having viscous and perturbative forces. The mean squared velocity has the Gaussian form with $<v^2(t)> \sim t$ in $\tau = 0$, in the case of a passive particle with both harmonic, viscous forces and that with viscous, perturbative forces.

In the case of a run-and-tumble particle, the mean squared displacement scales to $<x_\pm^2(t)> \sim t^3$ in $t \ll \tau$, while the mean squared velocity has a normal diffusive form with $<v_\pm^2(t)> \sim t$ in $t \gg \tau$ and $\tau = 0$. However, as has been studied so far, the mean squared displacement for a run-and-tumble particle generally scales as $<x^2(t)> \sim t^2$ in the small-time limit and as $<x^2(t)> \sim t$ in large-time limit. Compared with such result, our result of the mean squared velocity $<v_\pm^2(t)>$ is matched as the same value to time $\sim t$ at clockwise and counter-clockwise directions in $t \gg \tau$ and $\tau = 0$. We obtain for a passive particle having harmonic and viscous forces that $\mu_{2,2}$ is proportional to $t^5$ ($t^4$) in $t \ll \tau$ and $\tau = 0$ ($t \gg \tau$), consistent with our result. The moment $\mu_{2,2}$ for a passive particle with viscous and perturbative forces scales to $\sim t^6$ ($\sim t^5$) in $t \ll \tau$ ($\tau = 0$), consistent with our result. For a run-and-tumble particle, the moment $\mu_{2,2}^\pm$ scales as $\sim t^6$ in $t \ll \tau$, consistent with our result. Other values $\mu_{m,n}^\pm$ will be published elsewhere. Table 1 (Table 2 and Table 3) is summarized approximate values of the kurtosis, the correlation coefficient, and the moment in $t \ll \tau$, $t \gg \tau$ and $\tau = 0$ for a passive particle having harmonic and viscous forces (a passive particle with viscous and perturbative forces and a run-and-tumble particle with harmonic and viscous forces). We have obtained a simple solution from our model for a passive and a run-and-tumble particles. It is hoped from different forces contacted with the particle of systems that we will extend our model to novelly the generalized Langevin equation or the Fokker-Planck equation. The results obtained can be compared and analyzed with other theories, computer-simulations, and experiments [48-50].